\documentclass[10pt]{article}

\usepackage{amsmath}
\usepackage{amssymb}

\usepackage{graphicx}

\usepackage{cite}

\usepackage{color} 

\usepackage[labelfont=bf,labelsep=period,justification=raggedright]{caption}

\makeatletter
\renewcommand{\@biblabel}[1]{\quad#1.}
\makeatother

\date{}

\begin{document}

% Title must be 150 characters or less
\begin{flushleft}
{\Large
\textbf{Extrapolation of urn models via Poissonization: accurate measurements of the microbial unknown}
}
% Insert Author names, affiliations and corresponding author email.
\\
Manuel E. Lladser$^{1,\ast}$, 
Ra\'ul Gouet$^{2}$, 
Jens Reeder$^{3}$
\\
\bf{1} Department of Applied Mathematics, University of Colorado, Boulder, Colorado, USA
\\
\bf{2} Centro de Modelamiento Matem\'atico (CNRS UMI 2807), Universidad de Chile, Santiago, Chile
\\
\bf{3} Department of Chemistry and Biochemistry, University of Colorado, Boulder, Colorado, USA
\\
$\ast$ E-mail: manuel.lladser@colorado.edu
\end{flushleft}

% Please keep the abstract between 250 and 300 words
\section*{Abstract}
The availability of high-throughput parallel methods for sequencing microbial communities is increasing our knowledge of the microbial world at an unprecedented rate. Though most attention has focused on determining lower-bounds on the $\alpha$-diversity i.e. the total number of different species present in the environment, tight bounds on this quantity may be highly uncertain because a small fraction of the environment could be composed of a vast number of different species. To better assess what remains unknown, we propose instead to predict the fraction of the environment that belongs to unsampled classes. Modeling samples as draws with replacement of colored balls from an urn with an unknown composition, and under the sole assumption that there are still undiscovered species, we show that conditionally unbiased predictors and exact prediction intervals (of constant length in logarithmic scale) are possible for the fraction of the environment that belongs to unsampled classes. Our predictions are based on a Poissonization argument, which we have implemented in what we call the Embedding algorithm. In fixed i.e. non-randomized sample sizes, the algorithm leads to very accurate predictions on a sub-sample of the original sample. We quantify the effect of fixed sample sizes on our prediction intervals and test our methods and others found in the literature against simulated environments, which we devise taking into account datasets from a human-gut and -hand microbiota. Our methodology applies to any dataset that can be conceptualized as a sample with replacement from an urn. In particular, it could be applied, for example, to quantify the proportion of all the unseen solutions to a binding site problem in a random RNA pool, or to reassess the surveillance of a certain terrorist group, predicting the conditional probability that it deploys a new tactic in a next attack.

\vfill

Full paper available at:\\
\begin{center}
{\sc \hbox{http://www.plosone.org/article/info\%3Adoi\%2F10.1371\%2Fjournal.pone.0021105}}
\end{center}

\vfill

\end{document}